\documentclass[journal]{IEEEtran}
\usepackage{amsmath,amsfonts}
\usepackage{algorithm}
\usepackage{array}
\usepackage{subfig}
\usepackage{textcomp}
\usepackage{stfloats}
\usepackage{url}
\usepackage{verbatim}
\usepackage{graphicx}
\usepackage{cite}
\usepackage{xcolor}
\usepackage{multirow}
\usepackage[2]{editing}
\usepackage{amssymb}
\usepackage{algpseudocode}
\usepackage{graphicx}
\usepackage{multirow}
\usepackage{arydshln}
\usepackage{bm}
\usepackage{lipsum}

\begin{document}

\title{A Spatial Similarity-Guided Pilot Assignment and Access Point Selection for Cell-Free Massive MIMO Networks}

\author{Saeed~Mohammadzadeh, Kanapathippillai~Cumanan, Pei~Liu, and Hien~Quoc~Ngo

\vspace{-2em}
% \thanks{Pei Liu is with the School of Information Engineering, Wuhan University of Technology, Wuhan 430070, China (e-mail: pei.liu@ieee.org).}

}%,~\IEEEmembership{~IEEE,}}
        % <-this % stops a space
%\thanks{This paper was produced by the IEEE Publication Technology Group. They are in Piscataway, NJ.}% <-this % stops a space
%\thanks{Manuscript received April 19, 2021; revised August 16, 2021.}}


\maketitle

\begin{abstract}
% In this letter, we investigate pilot assignment and access point (AP) selection strategies for uplink cell-free massive multiple-input multiple-output (CF-mMIMO) systems. We propose a pilot assignment and AP clustering algorithm based on channel correlation to enhance the system's spectral efficiency (SE). The proposed pilot assignment strategy dynamically allocates pilots based on user channel correlation by assigning distinct pilots to highly correlated users. Then, the low-correlation AP selection algorithm is designed to enhance network performance by strategically selecting APs with minimal correlation. By reducing the correlation between APs chosen, the algorithm mitigates interference and improves spatial diversity while ensuring that users maintain strong and reliable channel conditions. This approach enhances SE and overall system robustness, particularly in dense deployment scenarios. Simulation results are presented to validate the effectiveness of the proposed algorithm in dynamic user demand scenarios.

This paper investigates pilot assignment and access point (AP) selection strategies for uplink cell-free massive multiple-input multiple-output (CF-mMIMO) systems. We propose channel similarity-aware pilot assignment (CAPA) and AP selection schemes to improve interference management and, consequently, spectral efficiency (SE). The pilot assignment strategy dynamically allocates pilot sequences by evaluating inter-user channel similarity, ensuring that users (UEs) with high channel similarity are assigned orthogonal pilots to mitigate pilot contamination. Subsequently, an AP selection algorithm is introduced that prioritizes the selection of low-correlation APs to reduce interference and enhance spatial diversity. This selection process maintains robust UE-AP links while minimizing inter-AP redundancy. The combined approach significantly improves SE, particularly in dense network deployments. Simulation results are provided to demonstrate the effectiveness of the proposed strategies under dynamic UE scenarios.

\end{abstract}


\vspace{-4mm}
\section{Introduction}
\IEEEPARstart{C}{ell}-free massive multiple-input multiple-output (CF-mMIMO) is emerging as a key technology for next-generation wireless networks\cite{mohammadzadeh2025pilot}. Unlike conventional cellular architectures, CF-mMIMO eliminates cell boundaries by deploying a large number of distributed access points (APs) that cooperatively serve all users (UEs) in a given area. By leveraging coherent signal processing, CF-mMIMO significantly enhances spectral efficiency (SE), coverage, and UE fairness. However, two challenges hinder its practical implementation: pilot contamination and access point (AP) selection.\\
\indent Pilot contamination occurs because the limited number of orthogonal pilots must be reused as the network grows. Since uplink pilots are crucial for channel estimation, pilot reuse introduces interference that distorts the channel estimation and degrades SE\cite{bjornson2017massive}. Mitigating this effect requires pilot assignment schemes that account for UE locations, AP–UE channel characteristics, and inter-UE interference. Pilot assignment in CF-mMIMO has been studied, beginning with simple random allocation and progressing to more structured approaches. Greedy refinement strategies have been proposed to improve assignments based on UE rate or location iteratively\cite{ngo2017cell, zhang2018location}. Optimization-based methods include max-cut formulations with weighted graph solutions\cite{zeng2021pilot}, and semidefinite relaxations for p-cut problems\cite{wang2024pilot}. Graph coloring\cite{Chen2019}, interference-aware grouping\cite{singh2023interference, aboulfotouh2025optimizing}, and spectral clustering\cite{zhang2023pilot} aim to suppress pilot contamination by exploiting spatial structure. Machine learning solutions have also emerged, including supervised pilot prediction from UE locations\cite{li2021scalable} and joint pilot/power optimization via multi-task learning\cite{khan2024joint}. Additional work includes weighted-graph contamination metrics\cite{hu2024graph}, and anchor-based clustering for fast pilot reuse\cite{zhao2025fast}.\\
\indent Another critical challenge is efficient AP selection \cite{mohammadzadeh2025association}. In large-scale CF-mMIMO systems, serving all UEs with all APs is neither practical nor energy-efficient\cite{bjornson2020scalable,liao2025optimizing}. Because only a subset of APs meaningfully contributes to each UE’s signal, numerous AP-selection and clustering strategies have been proposed. Early work includes graph-theoretic overlapping clusters\cite{dai2014uplink}, large-scale fading (LSF), and received signal strength (RSS)-based selection\cite{ shakya2020joint, duan2022pilot}. Distance-based clustering was introduced in\cite{lin2019user}. Structured solutions partition APs to reduce interference and contamination in\cite{chen2020structured}. Game-theoretic and distributed optimization schemes are proposed in\cite{wei2022user}. Segmented architectures impose non-overlapping AP clusters for scalable coordination\cite{wang2023clustered} while advanced optimization includes quantum bacterial foraging\cite{li2023access} and deep learning based in\cite{xu2025joint,banerjee2023access}. \\
% \subsection{Contribution}
\indent Motivated by the scalability limits of fully cooperative CF-mMIMO, where dense AP and UE deployments create excessive channel state information (CSI) overhead and severe pilot contamination, we develop channel similarity-aware pilot assignment (CAPA) and AP selection strategies to improve interference management.
First, we design a pilot assignment algorithm that exploits channel similarity, ensuring that UEs with high channel similarity are assigned orthogonal or minimally interfering pilots. This reduces pilot contamination and improves channel estimation quality, which is essential for reliable detection and power control. Second, we introduce a scalable AP selection method that leverages the low inter-AP correlation inherent in distributed CF-mMIMO. By selecting APs for each UE based on channel strength and spatial separation, the algorithm preserves strong links while avoiding interference from highly correlated APs, reducing both computational load and signaling overhead. These two approaches provide a scalable framework for interference-aware pilot allocation and AP-UE association in large CF-mMIMO networks.\\
\textit{Notations:} Bold lowercase and uppercase letters represent vectors and matrices. The expectation operator is $\mathbb{E}\{\cdot\}$, and $\mathcal{N}_{\mathbb{C}}(\bf{0},\mathbf{Y})$ denotes a circularly symmetric complex Gaussian distribution with zero mean $\bf{0}$ and covariance $\mathbf{Y}$. The set of complex numbers is denoted as  $\mathbb{C}$. The operators $(\cdot)^\mathrm{T}$, $(\cdot)^{-1}$, and $(\cdot)^\mathrm{H}$ denote transpose, inverse, and Hermitian, respectively. We denote the $N\times N$ identity matrix by $\mathbf{I}_N$. The indicator function is $\mathbb{1}(\cdot)$. The Euclidean norm and absolute value are denoted as $||\cdot||$ and $|\cdot|$, respectively.

% \begin{figure}[!]
% 	\centering
% 		\includegraphics[width=0.7\linewidth]{Cell-free.jpg}
%         % \vspace{-0.75em}
% 	\caption{Cell-free massive MIMO, user $k$ is served by a group of APs}
%      % \vspace{-0.5em}
% 	\label{NOMA figure}
% \end{figure}
\vspace{-2mm}
\section{System Model and Problem Formulation}
We consider a CF-mMIMO system with \(L\) APs, each equipped with \(N\) antennas, serving \(K\) single-antenna UEs. The APs are connected to a CPU via ideal fronthaul links. The channel between AP \(l\) and UE \(k\) modeled as correlated Rayleigh fading, 
\(\mathbf{h}_{kl} \sim \mathcal{N}_{\mathbb{C}}(\mathbf{0}, \mathbf{R}_{kl})\),
where \(\mathbf{R}_{kl}\) is the spatial covariance matrix and 
\(\beta_{kl} = \mathrm{tr}(\mathbf{R}_{kl})/N\) represents the large-scale fading. It is assumed that large-scale parameters are known at the CPU. To enable scalable operation, we define serving AP set for UE \(i\) is \(\mathcal{A}_i \subseteq \{1,\ldots,L\}\), meaning \(\mathbf{D}_{il} = \mathbf{I}_N\) for \(l \in \mathcal{A}_i\) and \(\mathbf{0}\) otherwise. The UEs served by AP \(l\) are collected in 
\(\mathcal{U}_l = \{i : \mathrm{tr}(\mathbf{D}_{il}) \ge 1\}\).
In each coherence block, \(\tau_p\) symbols are used for pilots and \(\tau_d = \tau_c - \tau_p\) for uplink payload data.
\vspace{-1em}
\subsection{Uplink Pilot Training and Channel Estimation}
We assume that pilot sequences are randomly assigned to the UEs, enabling a fully distributed operation. Each AP independently performs channel estimation using uplink pilots. A set of \(\tau_p\) mutually orthogonal pilot sequences \(\boldsymbol{\phi}_1,\ldots,\boldsymbol{\phi}_{\tau_p}\) is shared among the \(K\) UEs, with \(\tau_p < K\) and \(\|\boldsymbol{\phi}_t\|^2=\tau_p\). Let \(\mathcal{S}_t \subseteq \{1,\ldots,K\}\) denote the set of UEs using pilot \(t\). When these UEs transmit their pilots, the projection of the received pilot signal onto $\boldsymbol{\phi}_t$ normalized by $1/\sqrt{\tau_p}$ is given by\cite{bjornson2017massive}
\begin{equation} \label{RPS}
\mathbf{y}_{tl}^{\mathrm {p}} = \sum _{i \in \mathcal{S}_t} (\tau_p p_i)^{1/2} \mathbf {h}_{il}  + \mathbf{n}_{tl}^{\mathrm{p}}, 
\end{equation}
where $p_i$ is the transmit power of each UE and $\mathbf{n}_{tl}^{\mathrm{p}} \sim \mathcal{N}_{\mathbb{C}}({\bf 0}, \sigma^2 \mathbf{I}_N)$. The minimum mean-square error (MMSE) is utilized to estimate the channel of UE $k$ in $\mathcal{S}_t$ as \cite{bjornson2020scalable}
\begin{align}
    \hat{\mathbf{h}}_{kl} = (\tau_p p_i)^{1/2} \mathbf {R}_{kl} \mathbf{\Phi}_{tl}^{-1} \mathbf {y}_{tl}^{\mathrm {p}} \sim \mathcal{N}_{\mathbb{C}}({\bf 0}, \mathbf{B}_{kl}),
\end{align}
where $\mathbf{\Phi}_{tl} = \mathbb{E} \{ \mathbf{y}_{tl}^{\mathrm {p}} (\mathbf{y}_{tl}^{\mathrm {p}})^\mathrm{H}  \} =  \sum_{i \in \mathcal{S}_t} \tau_p p_i \mathbf{R}_{il} + \sigma^2 \mathbf{I}_N $ denotes the correlation matrix of the received pilot signal, and $\mathbf{B}_{kl} = \mathbb{E} \{ \hat{\mathbf{h}}_{kl} \hat{\mathbf{h}}_{kl}^\mathrm{H} \} = \tau_p p_k \mathbf{R}_{kl} \mathbf{\Phi}_{tl}^{-1} \mathbf{R}_{kl}$ while the estimation error is $\tilde{\mathbf{h}}_{kl} = \mathbf{h}_{kl} - \hat{\mathbf{h}}_{kl}$ and the correlation of the estimation error is defined as $\mathbf{C}_{kl} = \mathbb{E} \{ \tilde{\mathbf{h}}_{kl} \tilde{\mathbf{h}}_{kl}^\mathrm{H} \} =  \mathbf{R}_{kl} -\mathbf{B}_{kl} $.
\vspace{-1.5mm}
\subsection{Uplink Data Transmission }
The $l^\text{th}$ AP receives signal $\mathbf{y}_l^u \in \mathbb{C}^N$ in the uplink data transmission from all UEs as\vspace{-0.5em}
\begin{align}
    \mathbf{y}_l^u = \sum_{i=1}^K \mathbf{h}_{il} q_i + \mathbf{n}_l,
\end{align} 
where $q_i$ denotes the transmitted signal from UE $i$ and the receiver noise is depicted as $\mathbf{n}_l \sim \mathcal{N}_{\mathbb{C}} ({\bf 0}, \sigma^2 \mathbf{I}_N)$. 
For large-scale network deployment, we prefer offloading most computational tasks to the APs to avoid CPU overload. More specifically, every AP preprocesses its signal by computing local estimates of the data and then passes them to the CPU for final decoding. Although all APs receive signals from all UEs, only the APs in \(\mathcal{A}_{k}\) contribute to the detection of \(\mathrm{UE}\,k\) according to the AP selection policy. In this system, for each UE $k \in \mathcal{U}_l$, AP $l$ selects a combining vector $\mathbf{a}_{kl}$, to compute the local estimate of $q_k$ as $\hat{q}_{kl} = \mathbf{a}_{kl}^\mathrm{H} \mathbf{D}_{kl} \mathbf{y}_l^u $ and hence,
\vspace{-2mm}
\begin{align}
    \hat{q}_{kl} = \mathbf{a}_{kl}^\mathrm{H} \mathbf{D}_{kl} {\mathbf{h}}_{kl} q_k + \sum_{i \neq k}^K \mathbf{a}_{kl}^\mathrm{H} \mathbf{D}_{kl} {\mathbf{h}}_{il} q_i + \mathbf{a}_{kl}^\mathrm{H} \mathbf{D}_{kl} \mathbf{n}_l.
\end{align}
Thus, the achievable uplink SE is calculated using the use-and-then-forget method \cite{bjornson2017massive} as follows:
\begin{equation} 
\mathrm {SE}_{k}\!=\!\frac {\tau _{d}}{\tau _{c}} {\log _{2}\!\Bigg (\!{1 \!+\! \frac { p_{k} \mathbf{u^\mathrm{H}_k \mathbf{u}_k} }{ \sum \limits _{i=1}^{K} p_{i} \boldsymbol{\Xi}_{k i} \!-\!   p_{k} \mathbf{u}_k  \mathbf{u}_k^\mathrm{H}\!+\! \sigma^2 \boldsymbol{\Gamma}_k } \!\!}\Bigg)},
\vspace{-1.5em}
\end{equation}
where
\begin{align}
\mathbf{u}_k & =\left[\mathbb{E}\left\{\mathbf{a}_{k 1}^{\mathrm{H}} \mathbf{D}_{k 1} \mathbf{h}_{k 1}\right\}, \ldots, \mathbb{E}\left\{\mathbf{a}_{k L}^{\mathrm{H}} \mathbf{D}_{k L} \mathbf{h}_{k L}\right\}\right]^{\mathrm{T}}. \\
\boldsymbol{\Xi}_{k i} & =\left[\mathbb{E}\left\{\mathbf{a}_{k l}^{\mathrm{H}} \mathbf{D}_{k l} \mathbf{h}_{i l} \mathbf{h}_{i j}^{\mathrm{H}} \mathbf{D}_{k j} \mathbf{a}_{k j}\right\}: l, j=1, \ldots, L\right]. \\
\boldsymbol{\Gamma}_k & =\operatorname{diag}\left(\mathbb{E}\left\{\left\|\mathbf{D}_{k 1} \mathbf{a}_{k 1}\right\|^2\right\}, \ldots, \mathbb{E}\left\{\left\|\mathbf{D}_{k L} \mathbf{a}_{k L}\right\|^2\right\}\right).\\
\mathbf {a}_{kl} &= p_{k} \Big ({\sum \limits _{i \in \mathcal {U}_{l}} p_{i}  ({\hat { \mathbf {h}}_{il} \hat { \mathbf {h}}_{il}^{ {\scriptscriptstyle \mathrm {H}}} + \mathbf {C}_{il} }) + \sigma ^{2} \mathbf {I}_{N} }\Big)^{\!-1} \!\! \hat { \mathbf {h}}_{kl}. 
\end{align}
\vspace{-2em}
\section{Proposed pilot assignment and AP selection}
In this section, we propose the pilot assignments and AP clustering frameworks guided by channel similarity. The pilot assignment strategy dynamically allocates orthogonal or minimally interfering pilot sequences to UEs. Then, we introduce a low-correlation AP-selection algorithm that chooses, for each UE, a subset of APs with strong channel gains to that user while ensuring that the selected APs remain mutually low-correlated.
\vspace{-0.8em}
\subsection{Pilot Assignment} \label{Pilot assignement} 
In CF-mMIMO, effective pilot assignment plays a crucial role in limiting pilot contamination, which is a primary bottleneck for accurate channel estimation and coherent transmission. To address this challenge and prevent high-similarity UEs from being assigned the same pilot, we define a channel similarity metric that quantifies the similarity between the channel vectors of UE pairs. 
The basic pilot assignment algorithm is based on an idea that ensures that high channel similarity UEs do not share the same pilot. In this regard, the collective channel from all APs to UE $k$ is defined as ${\mathbf{h}}_k = [{\mathbf{h}}_{k1}^\mathrm{T}, \cdots, {\mathbf{h}}_{kL}^\mathrm{T}]^\mathrm{T} \in \mathbb{C}^{LN}$ and the similarity coefficient between the channels of two UEs, such as $k$ and $v$, is expressed as 
\vspace{-2mm}
\begin{align} \label{corr. coeff}
    \rho_{kv} = {\mathbf{h}_k^\mathrm{H}\mathbf{h}_v}
    /\left({ \| \mathbf{h}_k \| \| \mathbf{h}_v \|}\right), \quad \forall v,k \in \{ 1,2, \cdots, K\}.
\end{align}
However, a central difficulty in CF-mMIMO is identifying which UEs are intrinsically similar in their spatial signatures, because these similarities dictate pilot contamination, inter-UE interference, and ultimately the achievable performance. Any metric that relies on instantaneous channel realizations is fundamentally unreliable: the small-scale fading completely masks the underlying channel similarity coefficient spatial structure and provides no stable indicator of UE similarity. However, the pilot assignment based on only $\rho_{kv}$ from a single realization could fail since \(\rho_{kv}\) is random and strongly affects UE separability, pilot contamination, and interference in CF-mMIMO. 
To obtain a meaningful, statistically robust measure, we utilize the expected squared magnitude of the channel similarity coefficient
\vspace{-1mm}
\begin{align} \label{Expected}
\mathbb{E}\{|\rho_{kv}|^{2}\} =\mathbb{E}\left\{{|\mathbf{h}_{k}^\mathrm{H}\mathbf{h}_{v}|^{2}}/\left({\|\mathbf{h}_{k}\|^{2}\|\mathbf{h}_{v}\|^{2}}\right)\right\}. 
\end{align}
The quantity \(|\rho_{kv}|^{2}\) measures the power overlap between the spatial signatures of UEs \(k\) and \(v\), and crucially, depends only on their covariance matrices. In contrast, the instantaneous value of $\rho_{kv}$ is a random variable whose first moment is completely uninformative ($
\mathbb{E}\{\mathbf{h}_k^\mathrm{H}\mathbf{h}_v\} = 0$ )
which means any metric based on the sign or magnitude of the instantaneous inner product is dominated by noise rather than structure. Therefore, the second moment is the only statistically meaningful level at which spatial similarity can be evaluated by $\mathbb{E}\{|\mathbf{h}_{k}^\mathrm{H}\mathbf{h}_{v}|^{2}\}$. 
The quantity $|\mathbf{h}_{k}^\mathrm{H}\mathbf{h}_{v}|^{2}$ represents the instantaneous power overlap between the channel vectors of the two UEs, while its expectation (i) removes the randomness caused by small-scale fading, yielding a stable statistical measure of UE similarity, and (ii) reveals how the covariance structures of the channels overlap, which is the key factor determining inter-UE interference, pilot contamination, and UE separability in cell-free massive MIMO systems. Therefore, the expected second moment of the numerator constitutes the appropriate metric for analyzing and quantifying the average channel similarity between UEs.

To effectively calculate the expectation, we assume that $\mathbf{h}_k\sim\mathcal{CN}(\mathbf{0},\mathbf{R}_k)$,
 $\mathbf{h}_v\sim\mathcal{CN}(\mathbf{0},\mathbf{R}_v)$,
with $\mathbf{R}_k=\mathrm{diag}\!\left(\mathbf{R}_{k1},\dots,\mathbf{R}_{kL}\right)$, 
$\mathrm{tr}(\mathbf{R}_{kl}) = N\beta_{kl}$, 
and $\mathbf{R}_{k l}$ is directly related to the angular power spectrum as \cite{demir2021foundations}
\vspace{-1.5mm}
\begin{align}
\left[\mathbf{R}_{k l}\right]_{m, n}=\int P_{k l}(\theta) e^{j 2 \pi d(m-n) \sin (\theta) / \lambda} d \theta,
\end{align}
where $P_{k l}(\theta)$ is the power distribution over angle of arrivals.\\
\indent Let assume $X=\mathbf{h}_{k}^\mathrm{H}\mathbf{h}_{v}$ and we can write
\vspace{-1mm}
\begin{align} \label{Num approx}
\mathbb{E}\{|X|^{2}\}
&=\mathbb{E} \{\mathbf{h}_{v}^\mathrm{H}
(\mathbf{h}_{k}\mathbf{h}_{k}^\mathrm{H})\mathbf{h}_{v}\}
=\mathrm{tr}(\mathbf{R}_{k}\mathbf{R}_{v}).
\end{align}
Also, the norms satisfy that $\mathbb{E}\{ \|\mathbf{h}_{k}\|^{2}\}=\mathrm{tr}(\mathbf{R}_{k}), 
\mathbb{E}\{\|\mathbf{h}_{v}\|^{2}\}=\mathrm{tr}(\mathbf{R}_{v}),
$ and independence yields
\vspace{-3mm}
\begin{align} \label{Den approx}
\mathbb{E}\{\|\mathbf{h}_{k}\|^{2}\|\mathbf{h}_{v}\|^{2}\}
=\mathrm{tr}(\mathbf{R}_{k})\,\mathrm{tr}(\mathbf{R}_{v}).
\end{align}
By replacing \eqref{Num approx} and \eqref{Den approx} into \eqref{Expected}, the expected squared magnitude of the channel similarity can be written as
\vspace{-1.2mm}
\begin{align}
\mathbb{E}\{|\rho_{kv}|^{2} \}
\approx 
{\mathrm{tr}(\mathbf{R}_{k}\mathbf{R}_{v})}
/\left({\mathrm{tr}(\mathbf{R}_{k})\,\mathrm{tr}(\mathbf{R}_{v})}\right).
\end{align}
Because \(\mathbf{R}_{k}\) is block diagonal, $\mathrm{tr}(\mathbf{R}_{k}\mathbf{R}_{v})
= \sum_{l=1}^{L}\mathrm{tr}(\mathbf{R}_{kl}\mathbf{R}_{vl}),
\mathrm{tr}(\mathbf{R}_{k})=\sum_{l=1}^{L}\mathrm{tr}(\mathbf{R}_{kl})
$. Thus
\vspace{-1mm}
\begin{align} \label{final Expect}
\mathbb{E} \{|\rho_{kv}|^{2} \}
\approx 
\frac{
\sum_{l=1}^{L}\mathrm{tr}(\mathbf{R}_{kl}\mathbf{R}_{vl})
}{
\left(\sum_{l=1}^{L}\mathrm{tr}(\mathbf{R}_{kl})\right)
\left(\sum_{l=1}^{L}\mathrm{tr}(\mathbf{R}_{vl})\right)
}.
\end{align}
The quantity \(\mathrm{tr}(\mathbf{R}_{k}\mathbf{R}_{v})\) measures how much the UEs' spatial covariance eigenspaces overlap. A larger value indicates stronger similarity in angular distribution and thus higher average interference or pilot contamination. The normalization removes power dependence, producing a pure measure of spatial similarity determined solely by propagation statistics.

Using \eqref{final Expect}, the pilot assignment process is designed to maximize pilot reuse efficiency while preserving signal distinguishability among spatially proximate UEs. Initially, orthogonal pilots are allocated to a subset of UEs up to the available pilot pool size $\tau_p$. For the remaining UEs, pilot reuse is managed by identifying UEs with minimal channel similarity to those already assigned a given pilot. When a candidate UE has multiple pilot options, the one associated with the lowest aggregate similarity is selected. In situations where all pilots have been used, a fairness-based criterion is employed to select the least utilized pilot, ensuring balanced reuse across the network. 
This strategy achieves a twofold objective: it prevents the reuse of pilot sequences among UEs with strong channel similarity, and it promotes load balancing in pilot allocation to reduce excessive interference linked to overly reused pilots. By exploiting LSF information, which is readily available at the CPU and does not require frequent updates, the proposed method remains computationally efficient and scalable to large network deployments. The whole procedure is given in Algorithm 1. \vspace{-.7em}
\begin{algorithm}[t]
\caption{Pilot Assignment (Channel Similarity-Aware)}
\label{Algo pilot_assignment}
\begin{small} % reduce font size for entire algorithm (keeps formulas and comments readable)
\setlength{\itemsep}{0pt}
\setlength{\parskip}{0pt}
\begin{algorithmic}[1]
    \State \textbf{Input:} channel vectors $\mathbf{h}_k=[\mathbf{h}_{k1}^\mathrm{T},\ldots,\mathbf{h}_{kL}^\mathrm{T}]^\mathrm{T}$
    \State \textbf{Initialize:} Pilot array $\mathcal{L}_\text{UEs}=[t_k]_{k=1}^K$, set of assigned pilots $\mathcal{P}_a$, set of available pilots $\mathcal{P}_{\text{av}}$
    \State Compute \eqref{final Expect}
    \For{$k = 1$ to $K$}
        \If{$k \leq \tau_p$}
            \State $t_k \gets k$ \Comment{Assign unique pilot to the first $\tau_p$ UEs}
        \Else
            \State Most similar UEs with UE $k$: $v^* = \arg\max_{v < k} |\rho_{kv}|$
            \State Retrieve the pilot of the most similar UE: $ p_{v^*} = t_{v^*}$
            \State Available pilots: $ \mathcal{P}_{\text{av}} = \{1,\ldots,\tau_p\} \setminus \mathcal{P}_a$
            \If{$\mathcal{P}_{\text{av}} \neq \emptyset$}
                \State Assign pilot with  minimizes channel similarity to already assigned UEs:
                $t_k \gets \arg\min_{p \in \mathcal{P}_{\text{av}}} 
                \sum_{u = 1}^{k-1} |\rho_{ku}| \cdot \mathbb{1}(t_u = p)$
               \Else
                \State Compute current usage counts of each pilot:
                $c_i = \sum_{u=1}^{K} \mathbb{1}(t_u = i), \quad i=1,\ldots,\tau_p
                $
                \State Assign the least-used pilot (load balancing):
                $t_k \gets \arg\min_{i\in\{1,\ldots,\tau_p\}} c_i $       
            \EndIf
        \EndIf
    \EndFor
    \State \textbf{Output:} Pilot assignment array $\mathcal{L}_\text{UEs}=[t_k]_{k=1}^K$
\end{algorithmic}
\end{small}
\end{algorithm}
\vspace{-0.8mm}
\subsection{AP Selection}\label{AP selection}
In CF-mMIMO systems, efficient AP selection is critical for balancing signal quality and inter-AP interference. To this end, we propose a correlation-aware AP selection framework in Algorithm 2 that jointly considers spatial correlation among APs and the large-scale fading characteristics of UE–AP links. The central objective is to select a subset of APs for each UE such that \textit{(i)} selected APs exhibit sufficiently strong channel gains, and \textit{(ii)} mutual correlation among serving APs is minimized, thereby reducing redundant transmission and enhancing spatial diversity.
In this regard, the collective channel from all UEs to AP $l$ is defined as ${\mathbf{h}}_l = [{\mathbf{h}}_{1l}^\mathrm{T}, \cdots, {\mathbf{h}}_{Kl}^\mathrm{T}]^\mathrm{T}$ and the expected squared magnitude of the channel similarity coefficient of two APs, such as $l$ and $j$, is 
\vspace{-1mm}
\begin{align} \label{Expected APs}
\mathbb{E}\{|\eta_{lj}|^{2}\} =\mathbb{E} \left \{{|\mathbf{h}_{l}^\mathrm{H}\mathbf{h}_{j}|^{2}}/\left({\|\mathbf{h}_{l}\|^{2}\|\mathbf{h}_{j}\|^{2}}\right)\right \}, \!\!\!\quad \forall l,j \in L.
\end{align}
The algorithm proceeds by grouping APs into the set \( \mathcal{G} \) based on correlation, where an AP is included if its correlation with another AP is below a baseline gain level. This process ensures that the APs in each group exhibit low inter-AP correlation, reducing interference when serving the same UE. Following the grouping step, the algorithm iterates over each UE to determine which APs will be assigned. For each UE, it examines the APs within each low-correlation group and selects those that have strong links to the UE. Once the valid APs for the UE are identified, the algorithm filters out APs that have already been assigned in previous iterations to avoid redundancy. The final selection of APs for each UE is determined by updating the set \( \mathcal{A}_k \), which contains only those APs that satisfy both the low-correlation and strong-fading conditions. The algorithm outputs these sets for all UEs, providing an optimized AP allocation that balances signal strength and interference mitigation.

The approach effectively minimizes inter-AP interference by selecting APs with low correlation, which is critical for maximizing SE in CF-mMIMO systems. Additionally, the reliance on both correlation and fading thresholds ensures that UEs are served by APs with the best possible channel conditions while preventing unnecessary redundancy in AP assignment. This methodology improves network performance by optimizing SE, fairness, and overall system capacity, making it highly suitable for distributed massive MIMO architectures where APs collaboratively serve UEs in a dynamic and interference-limited environment.
\begin{algorithm}[!t]
\caption{AP Selection (Channel Similarity-Aware) }
\label{algo:low_corr_AP_selection_no_kappa}
\begin{small}
\setlength{\itemsep}{0pt}
\setlength{\parskip}{0pt}
\begin{algorithmic}[1]

    \State \textbf{Input:} $\beta_{kl}$, channel stack $\mathbf{h}_l=[\mathbf{h}_{1l}^\mathrm{T},\ldots,\mathbf{h}_{Kl}^\mathrm{T}]^\mathrm{T} \in \mathbb{C}^{NK}$
    \State \textbf{Initialize:} $\{\mathcal{A}_k\}_{k=1}^K=\emptyset$; set of less similar APs $\mathcal{G}$; \quad AP groups $\mathcal{AP}_{\mathcal{G}}$
    \State Compute channel similarity coefficient matrix using \eqref{Expected APs}
    \State Compute baseline gain threshold:
    $\alpha_{\mathrm{thre}}=\frac{1}{LNK}\sum_{l=1}^{L}\sum_{k=1}^{K}\|\mathbf{h}_{kl}\|^2$
    % \Comment{Identify low-correlated APs}
    \For{$l=1$ to $L$}
        \If{$|\eta_{lj}| < \alpha_{\mathrm{thre}}$}
            \State $\mathcal{G} \gets \{l\} \cup \mathcal{G}$
        \EndIf
    \EndFor
    % \Comment{User-specific AP selection}
    \For{$k=1$ to $K$}
        \For{$g=1$ to $|\mathcal{G}|$}
             \State $\mathcal{AP}_{\mathcal{G}} \gets \mathcal{G}[g]$ %\Comment{APs in correlation group $g$}
            \State Compute quality threshold for group $g$:
            $\beta^{(g)}_{\mathrm{thre}}
                =\frac{1}{|\mathcal{AP}_{\mathcal{G}}|}
                 \sum_{i \in \mathcal{AP}_{\mathcal{G}}} \beta_{ki}$
            \For{each $i \in \mathcal{AP}_{\mathcal{G}}$}
                \If{$\beta_{ki} \ge \beta^{(g)}_{\mathrm{thre}}$}
                    \State $\mathcal{V}_k \gets \mathcal{V}_k \cup \{i\}$ 
                    % \Comment{High-quality APs}
                \EndIf
            \EndFor

            \State $\mathcal{U}_k \gets \mathcal{AP}_{\mathcal{G}} \setminus \mathcal{V}_k$
            \State $\mathcal{A}_k \gets \mathcal{A}_k \cup \mathcal{U}_k$ 
            % \Comment{Select remaining APs for user $k$}

        \EndFor
    \EndFor

    \State \textbf{Output:} Selected AP sets $\{\mathcal{A}_k\}_{k=1}^K$

\end{algorithmic}
\end{small}
\end{algorithm}

The computational cost of Algorithm~1 is dominated by two parts. Computing pairwise UE correlations requires evaluating $\tfrac{K(K-1)}{2}$ inner products of $L$-dimensional vectors, giving a complexity of $\mathcal{O}(K^{2}L)$. The main loop over all $K$ UEs adds $\mathcal{O}(K^{2} + K\tau_p)$ from scanning correlated UEs, checking pilot availability, and updating counters. Thus, the total complexity is $\mathcal{O}(K^{2}L + K^{2} + K\tau_p)$, which is dominated by $\mathcal{O}(K^{2}L)$.
For Algorithm~2, the dominant cost is computing the AP--AP correlation matrix in Step~3, requiring $\mathcal{O}(L^{2}K)$. Step~4 (channel norms and thresholding) adds $\mathcal{O}(KL)$. The AP grouping introduces negligible additional cost after correlations are computed, and the UE-wise AP selection contributes $\mathcal{O}(KGL)$ with $G \leq L$. Hence, the overall complexity is $\mathcal{O}(L^{2}K + KL)$, dominated by $\mathcal{O}(L^{2}K)$ when $L \gg K$.
\vspace{-0.7em}
\section{Numerical Results}
\begin{figure}[!]
\centering
\subfloat[$L=100$, $N=4$ $\tau_p=10$ and number of UEs $K=30,100$]{\label{CDF} \includegraphics[width=0.45\textwidth]{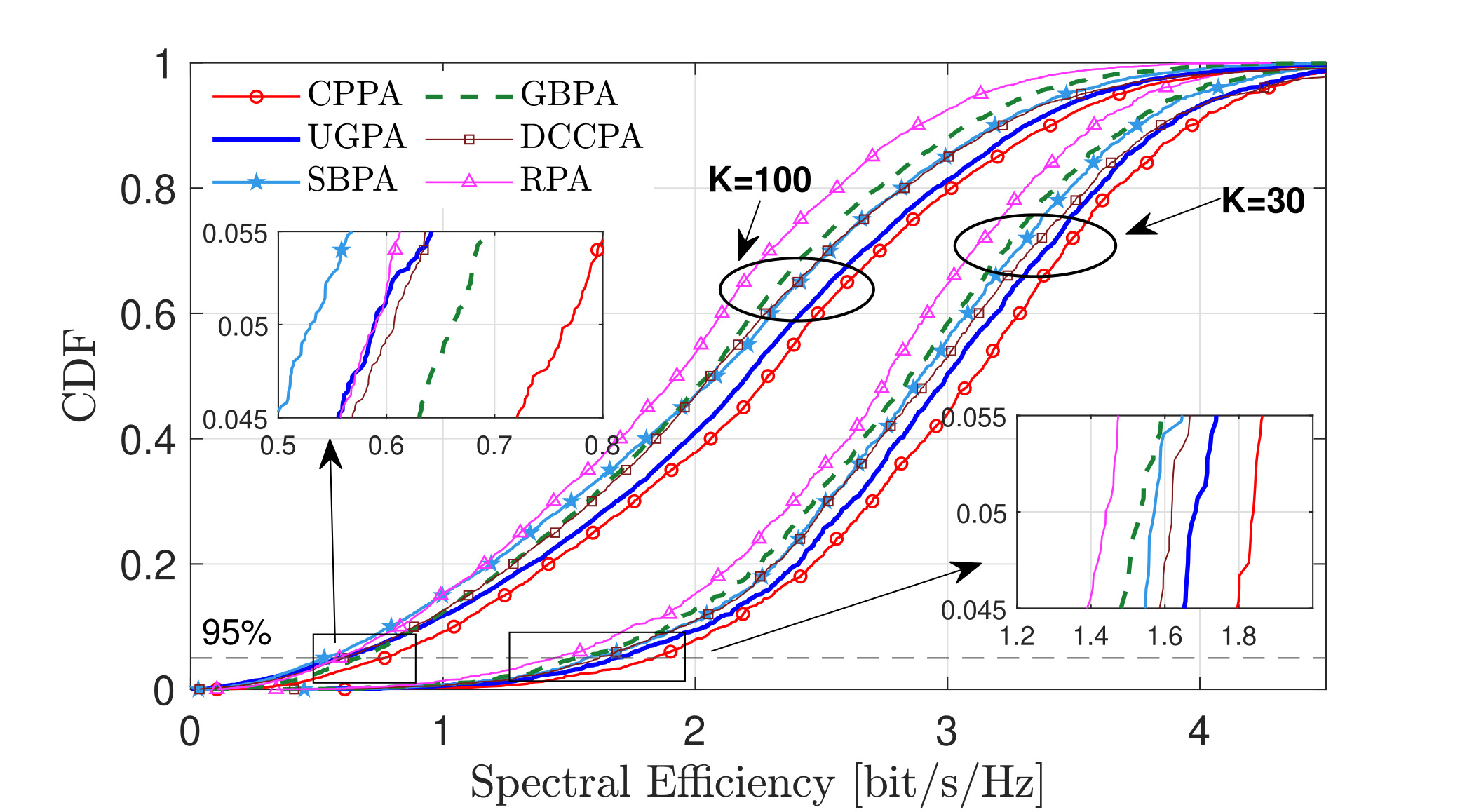}} 
 \vspace{-1.1em}
\subfloat[$K=40$, $\tau_p=10$, $N=1$, and $L=200,400$.]{\label{CDFAPs} \includegraphics[width=0.45\textwidth]{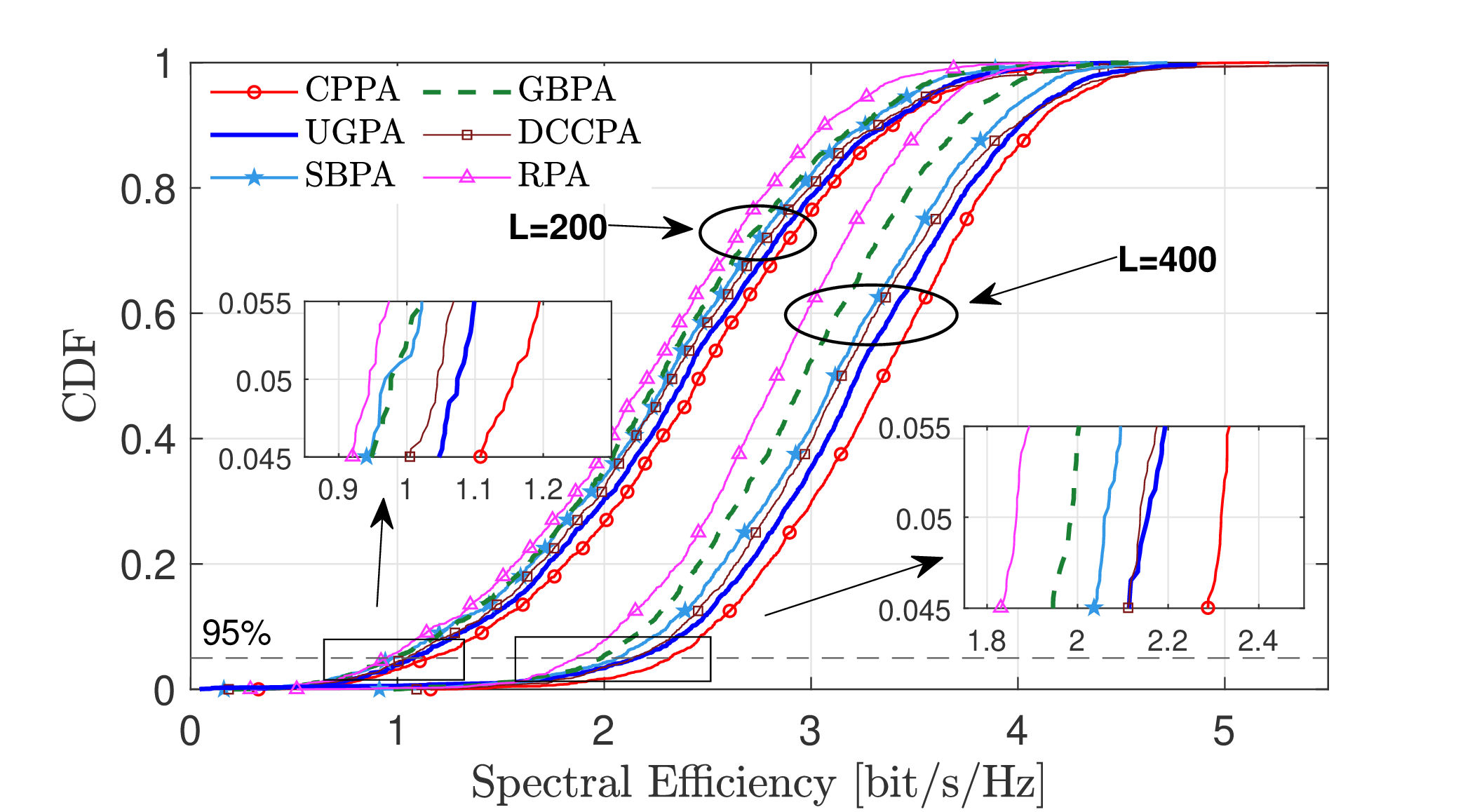}} 
\caption{SE per UE with different schemes}
\label{CDFBAR}
\end{figure}

\begin{figure}[!]
	\centering
	\includegraphics[width=0.45\textwidth]{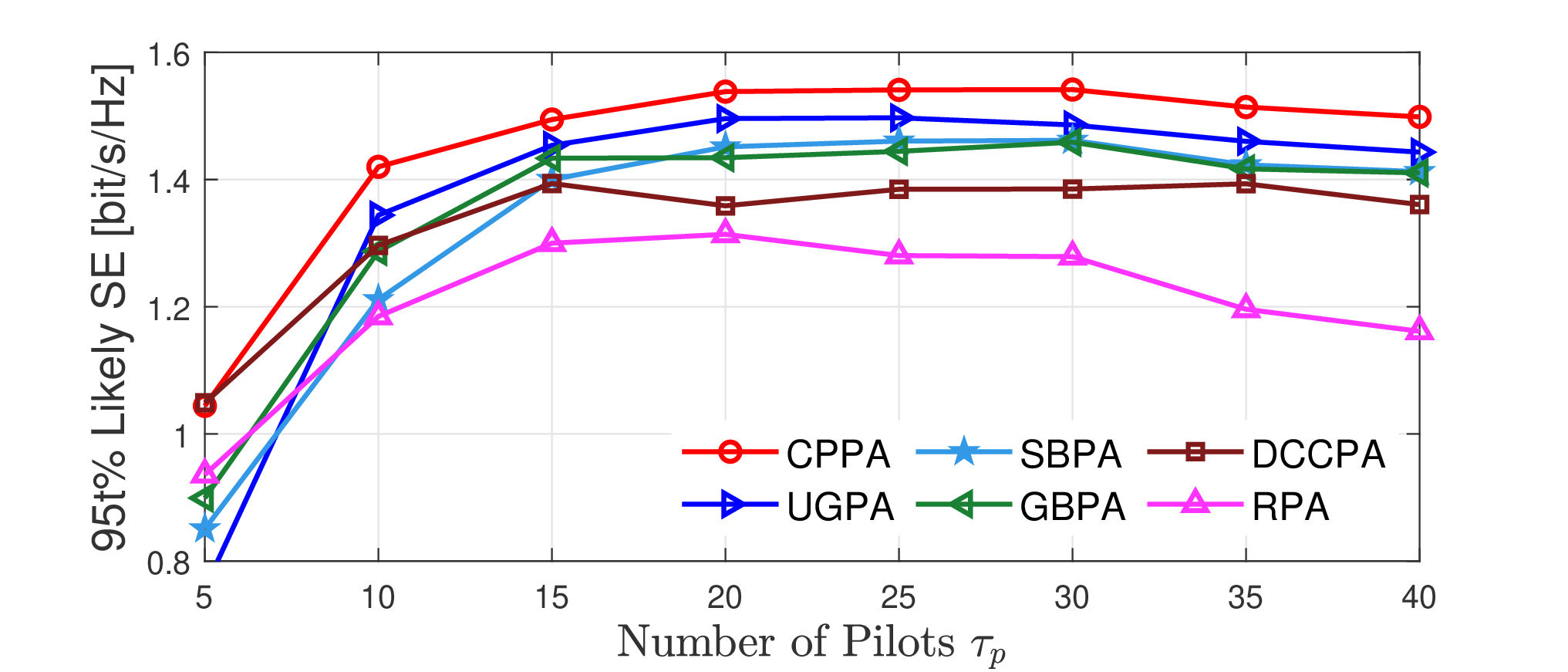}
    \vspace{-0.5em}
	\caption{95\% likely SE comparison with different number of pilot sequences $\tau_p$ for $L=100$, $N=4$, and $K=50$.}
 \label{tau}
\end{figure}

% \vspace{-1mm}
In this section, we evaluate the performance of the proposed pilot assignment and AP selection strategies, as outlined in Algorithms~1 and~2. We consider a system comprising $L = 100$  APs and $K$ UEs, which are independently and uniformly deployed within a $2\,\mathrm{km} \times 2\,\mathrm{km}$ square service area. The APs can be either arranged on a uniform square grid or randomly distributed across the coverage area. Each AP is equipped with a uniform linear array consisting of $N$ antennas, spaced at half-wavelength intervals. To emulate an infinitely large network and avoid boundary effects, we adopt the wrap-around technique. 
To model large-scale propagation effects such as pathloss and shadow fading, we utilize the 3GPP Urban Microcell scenario. The remaining system parameters align with those in~\cite{bjornson2020scalable}.
In simulation, we compare the proposed method with dynamic cooperation clustering (DCCPA) \cite{bjornson2020scalable}, UE-group-based pilot assignment (UGPA) \cite{chen2020structured}, graph-based (GBPA) pilot assignment \cite{Chen2019},  spectral-based (SBPA) pilot assignment \cite{zhang2023pilot}, and the random pilot assignment (RPA).\\
\indent Fig.~\ref{CDFBAR} compares the performance of the proposed pilot assignment and AP selection schemes under different UE and APs. In Fig.~\ref{CDF}, the CDF of the uplink SE per UE is shown for $K=30$ and $K=100$ UEs. When the number of UEs is relatively small, all methods except RPA achieve relatively high SE, but the proposed CPPA shows a clear improvement across the distribution. As the network becomes denser, $K=100$, the performance of all schemes degraded due to increased pilot interference; however, CPPA consistently maintains the highest SE. The inset plots highlight the behaviour near the 95\% region, demonstrating that CPPA preserves its advantage even in the tail where reliable performance matters most.\\
\indent The performance of the SE against different numbers of APs is presented in Fig.~\ref{CDFAPs}, where the number of antennas is fixed at $N=1$, and there is no spatial correlation. The superiority of the proposed scheme over other approaches is evident, particularly at 400 APs. It becomes clear that deploying many single-antenna APs is generally advantageous. The UEs with the weakest SE gain the most from a large number of APs, while the UEs with already strong SE achieve nearly the same performance even with fewer multi-antenna APs, due to the superior local interference suppression provided by those APs.\\
\indent Fig.~\ref{tau} shows the impact of the pilot-sequence length $\tau_p$. The proposed CPPA scheme consistently outperforms all benchmarks for every $\tau_p$, demonstrating the effectiveness of its pilot assignment and AP selection strategies. Increasing $\tau_p$ initially improves SE by reducing pilot contamination and enhancing channel estimation. However, beyond a certain point (e.g., $\tau_p = 20$), further increases degrade SE for RPA because more symbols are spent on pilots, leaving fewer for data transmission. This illustrates the trade-off between estimation accuracy and data throughput when selecting $\tau_p$.
\vspace{-0.5em}
\section{Conclusions}
This paper presented scalable pilot assignment and AP selection strategies. By leveraging channel similarity, the proposed pilot assignment reduces interference among UEs with similar channel characteristics. Simultaneously, the AP selection algorithm ensures that each UE is served by a subset of APs with strong, low-correlated links, thereby enhancing SE and reducing system complexity. Together, these methods enable efficient, interference-aware, and scalable approaches.
\vspace{-1em}

%\section*{Acknowledgments}
%This should be a simple paragraph before the References to thank those individuals and institutions who have supported your work on this article.
%\vspace{-3.5mm}
\bibliographystyle{IEEEtran}
\bibliography{deepref}

%\newpage
%\vfill

\end{document}